\documentclass[12pt]{article}
\usepackage{pdflscape}
\usepackage{calc}
\usepackage{color}
\usepackage{amsfonts}
\usepackage{latexsym}
\usepackage{placeins}
 \usepackage[dvips]{graphicx}
\usepackage{amssymb}
\usepackage{authblk}
\usepackage{amsmath}
\usepackage[cp1250]{inputenc}
\usepackage{soul}
\usepackage{multirow} 

\usepackage{setspace}

\newtheorem{lemma}{Lemma}
\newtheorem{corollary}{Corollary}
\newtheorem{theorem}{Theorem}

\newtheorem{definition}{Defintion}

\newcommand{\qed}{
\begin{flushright}
$\blacksquare$
\end{flushright}
}

\let\LaTeXtitle\title
\renewcommand{\title}[1]{\LaTeXtitle{\large\textsf{\textbf{#1}}}}

\title{Exact Solutions for Optimal Investment Strategies  and \\Indifference Prices under Non-Differentiable Preferences}
\date{}

\author[1]{M. Gaudenzi}
\author[2]{M.H. Vellekoop}
\affil[1]{
Dipartimento di Scienze Economiche e Statistiche,
University of Udine, Italy, 
{\tt marcellino.gaudenzi@uniud.it}}
\affil[2]{
Amsterdam School of Economics,
Faculty of Economics \& Business,
University of Amsterdam, The Netherlands,
{\tt m.h.vellekoop@uva.nl}}

\begin{document}

\maketitle 
\centerline{\today}

\newcommand{\mycomment}[1]{}
\newcommand{\be}{\begin{eqnarray*}}
\newcommand{\ee}{\end{eqnarray*}}
\newcommand{\mycomments}[1]{}
\newcommand{\quav}[2]{\left< #1,#2 \right>}
\newcommand{\sfrac}[2]{ {\textstyle\frac{#1}{#2}} }
\newcommand{\tu}{\tilde{u}}
\newcommand{\td}{\tilde{d}}
\renewcommand{\Re}{{\Bbb R}}
\renewcommand{\qed}{\hfill$\blacksquare$}
\newcommand{\one}[1]{ {\bf 1}_{ \{ #1 \}  } }
\bibliographystyle{plain}

\newcommand{\hsk}{\hat{\cal S}_k}
\newcommand{\hske}{\hat{\cal S}_{k+1}}

\begin{abstract}
We propose an algorithm to calculate the exact solution for utility optimization problems on finite state spaces under a class of non-differentiable preferences. We prove that optimal strategies must lie on a discrete grid in the plane, and this allows us to reduce the dimension of the problem and define a very efficient method to obtain those strategies. We also show how fast approximations for the value function can be obtained with an a priori specified error bound and we use these to replicate results for investment problems with a known closed-form solution. These results show the efficiency of our approach, which can then be used  to obtain numerical solutions for problems for which no explicit formulas are known.
\end{abstract}

\section{Introduction}

 One of the classical problems in mathematical finance concerns the optimal investment in risky assets by an investor who is risk averse. Explicit solutions for the trade-off between risk and return that characterize such problems were derived in the seminal work of Merton \cite{MertonOptimalConsumption}. His work showed that a certain combination of risk preferences and assumptions on the dynamics of asset prices leads to a stochastic control problem in continuous time which can be solved explicitly. In the most well-known example it is assumed that the risky asset prices are Geometric Brownian Motions and that risk tolerance is linear in wealth. In that case the optimal investment strategy turns out to be linear in wealth as well and an explicit formula can be derived for the proportion of wealth that is invested in the risky asset if the investor behaves optimally.

 This result has  been extended in many directions. Under the linear risk tolerance structure, better known as Constant Relative Risk Aversion (CRRA), more complicated asset dynamics can be treated.  One may still obtain relatively simple characterizations of optimal investment strategies when some of the parameters describing the dynamics of the risky asset vary over time in a deterministic way, for example. The resulting strategies are again linear in wealth but the coefficients will then vary over time. Kraft \cite{Kraft} has shown that this also holds when risky asset prices are generated by the stochastic volatility model introduced by Heston \cite{Heston}, which is generally considered to give a more realistic description of equity prices. 

 Another direction for generalizations also uses Geometric Brownian Motion process to describe asset prices, an assumption that is also known as Black-Scholes dynamics due to its use in the famous paper on option pricing \cite{BS}, but chooses different preferences. The class of utility functions known as Symmetrically Adjusted Hyperbolic Absolute Risk Aversion (SAHARA) also generates closed-form solutions. The optimal investment strategies are not linear and not even monotone in this case, since risk aversion is always positive but not always increasing when wealth levels become lower. Such preferences can therefore be used to describe the phenomenon where investors "gamble for resurrection", meaning that they may take more risky positions once their wealth levels become really low.

 For most utility functions and equity dynamics, no closed form solution can be derived for the optimal investment problem in continuous time. One therefore has to resort to numerical methods to generate suitable approximations to the optimal strategies. This makes calculations much more time consuming, which is in particular problematic when the optimal strategies are used as input for further calculations. 

 This is for example the case when one wants to determine what is known as the {\sl indifference price} for an asset or contingent claim which cannot be perfectly replicated using other assets in the market. Replication (in continuous time) means that a continuously updated portfolio can be defined which generates exactly the same payoff as a certain contingent claim. If this is the case, absence of arbitrage dictates that the price of the contingent claim is the same as the costs of setting up the initial portfolio which replicates it. But claims which cannot be replicated, i.e. claims which render the market {\sl incomplete} cannot be priced using such methods.

 An alternative definition for the selling price of such a claim states that the seller of the claim (who thus receives the price of the claim) should be indifferent, in terms of his or her expected utility, between selling the claim and receiving its price in compensation, or not selling the claim. Analogously, the buying price for such a claim should be chosen in such a way that paying this price at the initial time and receiving the payoffs of the claim afterwards, lead to exactly the same expected utility over the lifetime of the claim as not paying its price and not receiving its cashflows. This shows that these indifference pricing methods always lead to two different optimal investment problems that must be solved. Once involves optimal investment when the claim payoffs and initial buying or selling transactions are taken into consideration, while the other one involves optimal investment when there are no claims involved. Requiring the solutions to those two problems to be the same then implicitly defines what the price of a claim, i.e. the {\sl indifference price} should be.

 In practice, there are not that many indifferent pricing problems that can be solved explicitly. One therefore often has to rely on numerical approximations which are based dynamics in discrete time.
 In this paper we will show that exact solutions can be found for optimal investment and indifference pricing problems in discrete time if risk preferences are characterized by a class of utility functions which are piecewise linear. 
We require asset prices to be Markovian on a finite state space, and we take the binomial model of Cox, Ross and Rubinstein \cite{CoxRosRubinstein} as canonical example. We show that our class of utility functions is defined has certain properties which are inherited if they are propagated backwards in time under the dynamic programming equations which characterize optimal policies. As a result, we can prove that efficient algorithms exist which generate the exact solution for those policies and the associated value functions or indifference prices.

Our technique is based on a grid constructed in the $(w,b)$ plane, where $w$ is the total wealth and $b$ is  the wealth invested in risky assets. Such a grid consists of two sets of parallel lines with different slopes and we prove that the optimal strategy must always lie on this grid. This property allows us to define a method, which to the best of our knowledge has not been proposed before, to determine the optimal strategy in a very efficient way. When more risk factors are involved, such as a stochastic volatility component, we can still reduce the analysis of such more complicated problems to the design of a suitable grid on which optimal strategies must lie, and this testifies to the flexibility of the method we propose.

 We also show that very efficient algorithms can be defined which generate approximations to the exact optimal investment policies and value functions if one is willing to allow small errors. These errors can be guaranteed to stay smaller than an a priori specified error tolerance. The use of functions that are piecewise linear and thus characterized by their {\sl singular points} to determine exact and approximate solutions has been used before in the context of option pricing, see \cite{GaudenziLepellereZanette}.  

 To illustrate the working of the algorithm, we define discretized versions of the equity models mentioned above and approximate the corresponding risk preferences using members of our specific class of utility functions. This allows us to reproduce the optimal strategies derived for these very special cases with a very high accuracy. However, we believe that our method is particularly useful in cases where no closed-form strategies for the continuous-time version of the investment problem are known, or when one wants to study  the optimal behaviour of investors with non-differentiable preferences in a discrete time setting.

 The structure of the paper is as follows. In the following section we define the asset price dynamics and non-differentiable preferences that together characterize our optimal investment problem. In Section \ref{sec:dpe} we prove the main results of the paper and Section \ref{sec:numericalexamples} then applies our algorithm in a number of illustrative cases. We draw conclusions and discuss possible extensions of our method in the last Section.

\section{The Optimization Problem}

 In this section we specify our model and introduce our main assumptions, which concern the behaviour of risky assets and the risk preferences of the investor. Asset price dynamics are considered on a finite horizon in discrete time and must be Markovian. Asset prices are restricted to lie on a lattice in which every price has two possible successor price values one time step later\footnote{This binomial assumption could be extended to, for example, a trinomial specification but since every trinomial step can be described by two recombining binomial steps we restrict ourselves to the binomial case.}.  Investors' preferences must correspond to a utility function which is a member of a particular set of functions, which we call class $\cal H$. 

\subsection{Optimal Investment in Risky Assets}

We define for a given maturity $T>0$ and tree size $n\in {\Bbb N}\setminus\{0\}$ a binomial tree $${\cal T}=\displaystyle\bigcup_{m=0}^n {\cal T}_m,\qquad {\cal T}_m=\bigcup_{k=0}^{N(m)}\{(m\Delta t, S_k^m)\}, 
$$ with $N(m)\in {\Bbb N}$ the number of possible asset values at timestep $m$,  $S_k^m>0$, and functions $u:{\cal T}\to{\Bbb R}$ and $d:{\cal T}\to{\Bbb R}$ which describe possible transitions of the risky asset in terms of rates of return, and which should satisfy
$$
((m+1)\Delta t,\,  u(m\Delta t,S_k^m)\cdot S_k^m)\in {\cal T}_{m+1},\quad
((m+1)\Delta t,\, d(m\Delta t,S_k^m)\cdot S_k^m)\in {\cal T}_{m+1},
$$ 
 for all $0\leq m\leq n$ and $0\leq k\leq N(m)$.
We take  $\Delta t=T/n$ and define the riskfree return $R(t)=e^{r(t)\Delta t}$ and require that
$u(t,S)>R(t)>d(t,S)$ for all $(t,S)\in{\cal T}$. The probability that a transition from $(t,S)$ to $(t+\Delta t,u(t,S)S)$ will take place is denoted by $p(t,S)$ and the transition to $(t+\Delta t,d(t,S)S)$ thus has probability $1-p(t,S)$. The mean rate of return for the riskfree rate $\mu(t,S)$ is defined by $$p(t,S)u(t,S)+(1-p(t,S))d(t,S)=e^{\mu(t,S)\Delta t}$$ and the riskneutral probabilities $q(t,S)$ for this economy are the ones which satisfy $$q(t,S)u(t,S)+(1-q(t,S))d(t,S)=R(t)=e^{r(t)\Delta t}$$ so 
$$
p(t,S) \ = \  \sfrac{e^{\mu(t,S)\Delta t}-d(t,S)}{u(t,S)-d(t,S)},\qquad\qquad
q(t,S) \ = \  \sfrac{R(t)-d(t,S)}{u(t,S)-d(t,S)}.
$$
The functions $N$, $u$, $d$, $r$ and $\mu$ need to be specified in our setup, and then $p$ and $q$ follow from the previous equations. 
The well know standard binomial tree model introduced by Cox, Ross and Rubinstein \cite{CoxRosRubinstein} corresponds to choosing $$N(m)=m, \quad u(t,S)=\sfrac{1}{d(t,S)}=e^{\sigma\sqrt{\Delta t}}, \quad r(t)=r,\quad \mu(t,S)=r+\sfrac12 \sigma^2,$$ in this specification.

\bigskip

In this economy we will aim to maximize expected utility over a a set of allowed investment strategies $\phi_t(X_t)$ which must be in the set $\Phi$ of all processes which are a function of times $t=m\Delta t$ on the tree ${\cal T}$ and a given vector state process $X_t=(S_t,W_t,{\cal Y}_t)\in {\Bbb R}^l$ (with $l\geq 2$) which may contain, besides the current stock price $S_t$ and current wealth $W_t$, other information known at time $t$ which is collected in the vector ${\cal Y}_t$. This vector may be empty or contain additional observable information; we will treat, for example, the case of stochastic volatility or an untradeable process in later sections.
The functions $\phi\in\Phi$ map $\{0,\Delta t,...,n\Delta t:=T\}\times {\Bbb R}^l$ to $\Bbb R$ and have the interpretation of the value of the wealth that is invested in the risky asset $S$.

For a given utility function $U$ on the real line (i.e. a function which is increasing and concave) we define the optimization problem 
\begin{eqnarray}\label{eq:1}
\max_{\phi\in\Phi}{\Bbb E}[U(W_T^\phi)]
\end{eqnarray}
subject to 
\begin{eqnarray}\label{eq:Wdynamics}
W_{t+\Delta t}^\phi&=& \phi_t(X_t)\frac{S_{t+\Delta t}}{S_t} + (W_t^\phi-\phi_t(X_t))\, R(t)
\end{eqnarray}
where $(t,S_t)\in {\cal T}$ is the Markov chain we defined above on our tree $\cal T$,
and $W_0^\phi=w_0$ with $w_0\in\Re$ given.

\subsection{Non-differentiable Preferences}

 We will always assume that the utility function $U$ of our investment problem \eqref{eq:1} is in a class  $\mathcal{H}$ of functions on the real line ($U\in{\cal H})$ defined below.

\begin{definition}\label{def:1}
The class  $\mathcal{H}$ consists of functions $f:\Re \to \Re$ such that\\ 1. $f$ is piecewise linear with a finite number of  points where it is not differentiable,\\
	2. $f$ is concave, and \\
	3.  there exists an $\bar{x}\in\Re$  such that $f(x)$ is constant for $x\geq \bar{x}$.
\end{definition}
It immediately follows that any $f\in {\cal H}$ is continuous on its entire domain $\Re$ and that both its right-hand side derivative $f^{\prime +}(x)=\lim_{y\downarrow x}f'(x)$ and lefthand-side derivative
$f^{\prime -}(x)=\lim_{y\uparrow x}f'(x)$ exist. Since the derivative equals zero for large enough values and can only decrease, functions in ${\cal H}$ are increasing.  By concavity, the right-hand side derivative must always be equal or smaller than the left-hand side derivative and we call the finite set of unique, (increasingly) ordered points where these two derivatives are unequal, i.e. where $f^{\prime +}(x)<f^{\prime -}(x)$, the set of {\sl singular values}, denoted by $(x_k)_{k=1,...,N}$, see \cite{GaudenziLepellereZanette}. A function $f\in{\cal H}$ is uniquely characterized by its singular values, the values  $(f_k)_{k=1,...,N}$ in its singular values and its left-hand side derivative at the smallest singular value, $f^{\prime -}(x_1)$. 

\begin{lemma}\label{lemma0}
Assume $f_1,f_2\in{\cal H}$ and let $c_i\in\Re$ for $i=1...6$.  
\begin{itemize}
\item[(i)] For $c_1,c_2\geq 0$ we have that $c_1f_1+c_2f_2$ are in ${\cal H}$  and so is $x\to f_1(-x)+xf_1^{'-}(x_1)$ with $x_1$ the first singular value of $f_1$.
\item[(ii)] Assume that $c_1,c_2>0$ and $c_3c_5<0$ and define $f(x)= c_1f_1(c_3x+c_4)+c_2f_2(c_5x+c_6)$. The set $F=\textstyle\arg\max_{x\in\Re} f(x)$  is a non-empty compact interval in $\Re$ (which may consist of a single point).
\end{itemize}
\end{lemma}
{\bf Proof.}\ The first statement of (i) is immediate. The function $x\to f_1(-x)$ is concave if $f_1$ is, and adding $xf_1^{'-}(x_1)$ makes the function constant for $x$ large enough. For (ii) we first notice that $f'(x)$ goes to $-\infty$ for $x\to +\infty$ and for $x\to -\infty$ and $f$ is continuous on $\Re$ which implies that it attains a maximal value $f_{max}$ on $\Re$. The set $F=f^{-1}(\{f_{max}\})$ is bounded and closed.
Since $f$ is concave this set must be an interval. \qed

\section{Main Results}\label{sec:dpe}

 The combination of the asset price properties and risk preferences defined above will now allow us to prove a number of results which lead to an explicit characterization of the optimal investment policies.

\subsection{Dynamic Programming}\label{subsec:dpe}

The Dynamic Programming Principle (See for example \cite{bookBertsekasShreve}) gives a backward recursion for the value function $\tilde{V}_{t,X}$ for our optimization problem in the state $X$ at a time $t$ on the tree. 
For the problems we consider in this paper, we will be able to represent the value function
$$
\tilde{V}_{t,X_t} = \max_{\phi_t,\phi_{t+\Delta t},...,\phi_T} {\Bbb E}[\, U(W_T^\phi)\mid X_t\, ]
$$
using a function of wealth which is in ${\cal H}$ for every value of the rest of the state vector. In this section, for example, we use a state $X_t=(S_t,W_t)$ and we will write $V_{t,S_t}(W_t)$ for its value in such states. 
We use the notation $\beta_{t,S_t}(W_t)$ for the smallest value which makes the strategy $\phi$ defined by $\phi_t(X_t)=\beta_{t,S_t}(W_t)$ optimal. 

This implies, by \eqref{eq:Wdynamics} and the the Dynamic Programming Principle, that
\be
V_{t,S}(w)&=&\max_{b\in \Re} H_{t,S}(w,b), \qquad\quad B_{t,S}(w) \ =\ \underset{b \in \Re}{\arg\max} \ H_{t,S}(w,b),\\ \ \beta_{t,S}(w)&=& \min \{b:b\in B_{t,S}(w)\},
\ee
with
\begin{eqnarray}
H_{t,S}(w,b)&=&\nonumber
R(t)^{-1}\ \left[ \  
p(t,S)V_{t+\Delta t,u(t,S)S}(wR(t)+b(u(t,S)-R(t))) \ +\ \right. \\
&&  \qquad\quad \left.   (1-p(t,S))V_{t+\Delta t,d(t,S)S}(wR(t)+b(d(t,S)-R(t)))
\ \right].\label{eq:2long}
\end{eqnarray}
The functions $H_{t,S}:\Re^2\to\Re$ and $V_{t,S}: \Re \to \Re$  are defined for every $(t,S)\in {\cal T}$. We must have that $V_{T,S}=U$ for every $(T,S)\in {\cal T}_n$ which means that the value function is of class $\cal H$ in all nodes ${\cal T}_n$ at the final time $T$. We will show in this section that this property is inherited by the value function in all earlier states at all earlier times. We do this by assuming that for any given $(t,S)$, both $V_{t+\Delta t,u(t,S)S}$ and $V_{t+\Delta t,d(t,S)S}$ are in this class, and by proving that the same holds for $V_{t,S}$. To lighten notation we will from now on suppress the dependency on $(t,S)$ for the functions $u$ and $d$ and the time dependence of $R$ when no confusion can arise. The expression for $H_{t,S}$ then becomes, for example,
\begin{eqnarray}
H_{t,S}(w,b)&=&\nonumber
R^{-1}\ \left[ \  
pV_{t+\Delta t,uS}(wR+b(u-R)) \ +\ \right. \\
&&  \qquad\quad \left.   (1-p)V_{t+\Delta t,dS}(wR+b(d-R))
\ \right].\label{eq:2}
\end{eqnarray}

\bigskip

To prove that the properties of value functions in class $\cal H$ are preserved when applying the dynamic programming equations, we denote by $x_i^u$, $i=1...N_u$ and $x_j^d$, $j=1...N_d$ the singular values of $V_{t+\Delta t,uS}$ and $V_{t+\Delta t,dS}$, respectively. We also set $x_0^u=x_0^d=-\infty$, $x_{N_u+1}^u=x_{N_d+1}^d=+\infty$.
We see from \eqref{eq:2} that it will be useful to characterize the points $(w,b)$ where $wR+b(u-R)$ equals a singular value for $V_{t+\Delta t,uS}$ or where 
$wR-b(R-D)$ equals a singular value for $V_{t+\Delta t,dS}$.  
We will call this subset of $\Re^2$ the \textit{grid} $G$.
We thus define on ${\cal N}=\{(i,j):1\leq i\leq N_u,\, 1\leq j\leq ...N_d\}$ the sets $$D_{ij}=\{(w,b) \in \Re^2: \ x_i^u<wR+b(u-R)<x_{i+1}^u, \ x_j^d<wR-b(R-d)<x_{j+1}^d \},$$
and the grid $G
=\bigcup_{i=1}^{N_u}C_i^u \ \cup \ \bigcup_{j=1}^{N_d}C_j^d$
for
\begin{eqnarray}
C_i^u&=&\{(w,b)\in \Re^2: \ wR+b(u-R)=x_i^u\}, \label{eq:cu}\\
C_j^d&=&\{(w,b)\in \Re^2: \ wR-b(R-d)=x_j^d\}. \label{eq:cd}
\end{eqnarray}
To give unique coordinates to every point of the grid we also define:
\begin{eqnarray*}
L_{ij}^u&=&\{(w,b) \in \Re^2: \ wR+b(u-R)=x_i^u,  \ x_j^d \leq wR+b(d-R)<x_{j+1}^d \}\\ 
L_{ij}^d&=& \{(w,b) \in \Re^2: \ wR+b(d-R)=x_j^u,  \ x_i^u \leq wR+b(d-R)<x_{i+1}^u \} \\
L_{ij}&=&L_{ij}^d \cup L_{ij}^u.
\end{eqnarray*}

In the next lemmas, we take $(t,S)$ to be any point in ${\cal T}$.

\begin{figure}[t!]
\begin{center}
\includegraphics[width=0.8\textwidth]{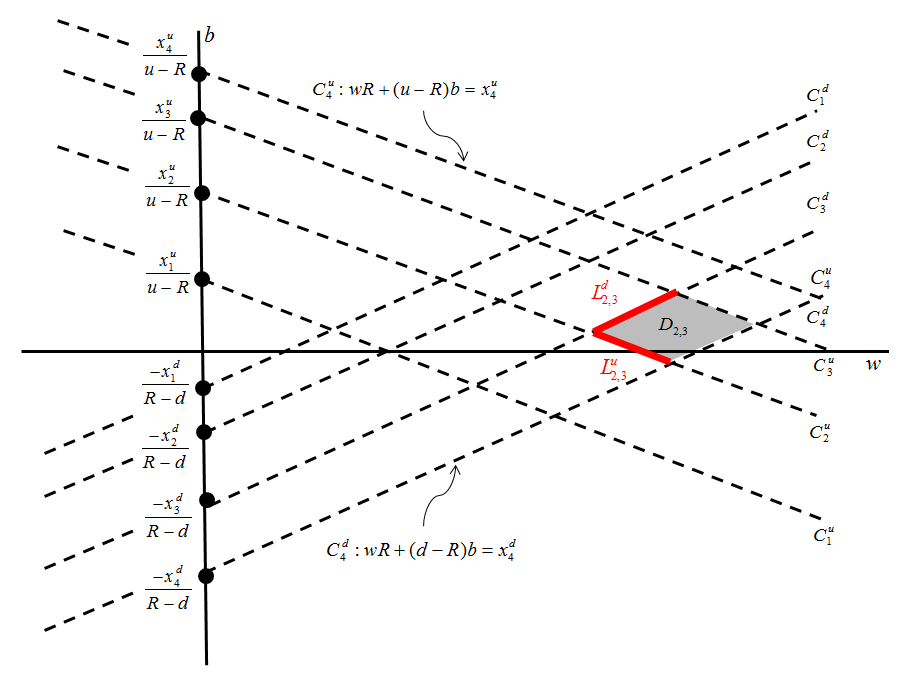}
\caption{The grid. Notice that the singular values have been taken positive here to show a clearer picture, but they may of course be negative as well.}
\end{center}
\end{figure}

\begin{lemma}\label{lemma1} For every $w \in \Re$, $b \to H_{t,S}(w,b)$ is concave, and for every $b \in \Re$, $w \to H_{t,S}(w,b)$ is concave. For both these functions the derivative strictly decreases at an intersection with the grid, i.e. when $(w,b)\in G$. 
\end{lemma}
{{\bf Proof.}\ In both cases the functions are sum of compositions of a concave piecewise linear  function and a linear function. Such a composition returns a concave piecewise linear function. An intersection with the grid means that $(w,b)$ is in $C_i^u$ for certain $i$ or in $C_j^d$ for certain $j$ (or both). This means that the derivative of $V_{t+\Delta t,uS}(wR+b(u-R))$ or $V_{t+\Delta t,dS}(wR-b(R-d))$ strictly decreases by the definition of singular value, while the other derivative must either stay constant or decrease as well. \qed

\bigskip

In the next lemma we show that if $b$ is a point which maximizes $H_{t,S}(w,b)$ for a certain $w$ and $b$ does not lie on the grid, then the function $H_{t,S}$ is constant in $b$ in the closure $\overline{D_{ij}}$ of the parallelogram  $D_{ij}$ to which belongs. Moreover, for this $w$ no other optimal points $b$ exist outside $\overline{D_{ij}}$. 

\begin{lemma}\label{lemma2}
If $b\in (B_{t,S}(w)\cap D_{ij})$ then $H_{t,S}(w,b)=H_{t,S}(w,b_1)>H_{t,S}(w,b_2)$ for all $(w,b_1)\in \overline{D_{ij}}$ and all  $(w,b_2)\not\in \overline{D_{ij}}$. 
\end{lemma}
{\bf Proof.}\ The function $H_{t,S}$ is linear on $D_{ij}$, since it is the sum of
two compositions of linear functions with (fixed) linear functions as long as we stay in $D_{ij}$. If $b$ is an internal maximum of the function $b\to H_{t,S}(w,b)$ for the fixed value of $w$, the function must be constant for all $b_1$ such that $(w,b_1)\in \overline{D_{ij}}$. Since the derivative of $b\to H_{t,S}(w,b)$ strictly decreases when $(w,b)\in G$ and using part (ii) of Lemma \ref{lemma0} shows that there can be no optimal points $(w,b_2)$ outside $\overline{D_{ij}}$.
 \qed

\bigskip

\begin{lemma}\label{lemma2a}
The optimal trajectory ${\cal B}=\{(w,\beta(w))|w\in\Re\}$ is a subset of $G$ and contains $L^u_{N_u,N_d}$. 
\end{lemma}

{\bf Proof.}\ By Lemma \ref{lemma2}, if an optimal point $(w,\beta(w))$ is not on $G$ but in a parallelogram $D_{ij}$, then all points of $\overline{D_{ij}}$ which have the same value of $w$ are optimal and no points outside $\overline{D_{ij}}$ are optimal for that $w$. Since $\beta(w)$ is defined as the minimum of all optimal points for a given $w$, we must have that $(w,\beta(w))\in G$.

The parallelogram $D_{N_u,N_d}$ contains wealth values beyond the last intersection point on the grid $G$, so the function $H_{t,S}$ is constant on $D_{N_u,N_d}$. This implies, by Lemma \ref{lemma1}, that every point in $D_{N_u,N_d}$ is optimal. For values of $w$ beyond the last intersection point we may thus conclude that that $(w,\beta(w))\in L^u_{N_u,N_d}$.
\qed

\bigskip

\begin{lemma}\label{lemma3}
For all $(t,S)\in {\cal T}$, the functions $\beta_{t,S}$ and  $V_{t,S}$ are continuous on their domain $\Re$. 
\end{lemma}
{\bf Proof.}\ Let $w_0 \in \Re$. By  Lemma \ref{lemma2}, $(w_0,\beta(w_0))$ must\footnote{We use the notation $\beta$ for the function $\beta_{t,S}$ here to lighten notation.} be on the grid $G$. Suppose that there exists an $\epsilon>0$ such that for all $b
\in [\beta(w_0),\, \beta(w_0)+\epsilon]$, $H_{t,S}(w_0,b)$ is constant and $(w_0,b)\in D_{ij}$. Since $H_{t,S}(w,b)$ is linear in $(w,b)$ on $D_{ij}$, the function $b\to H_{t,S}(w_1,b)$ must also be constant for all values $w_1$ and $b$ such that $(w_1,b)\in D_{ij}$, and it must attain smaller values than this constant for points $(w_1,b)$ outside $\overline{D_{ij}}$. By the definition of $\beta(w)$ as the minimum of all possible optimal points for a fixed $w$, we conclude that all points on the two lower sides of $D_{ij}$ are optimal points $(w,\beta(w))$.  This proves the continuity of $\beta$ at the point $w_0$.

Assume now that $\beta(w_0)$ is the unique maximum of $b \to H_{t,S}(w_0,b)$. Let $(w_n)_{n\geq 1}$ be a sequence converging to $w_0$ and set $\beta_n=\beta(w_n)$.
We know that the optimal points must be on the grid and that the grid is formed by a finite number of lines which are not vertical. Therefore the sequence $\beta_n$ must be bounded. Passing to a subsequence if necessary we can assume that $\beta_n$ converges to a value $b_0 \in \Re$, so  $(w_n, \beta_n)$ converges to $(w_0,b_0)$. To prove continuity of $\beta_{t,S}$ in $w_0$, we must show that $b_0=\beta(w_0)$.

  Since $w\to H_{t,S}(w,\beta(w_0))$ is continuous, there must exist for any given $\epsilon>0$ a $\delta>0$ such that $H_{t,S}(w,\beta(w_0))>H_{t,S}(w_0,\beta(w_0))-\epsilon$ for $w\in[w_0-\delta, w_0+\delta] 
$. Assume $w_n \in [w_0-\delta, w_0+\delta]$ for $n\geq n_\delta$; for such $n$ we have, by the optimality of $(w_n,\beta_n)$, that $H_{t,S}(w_n,\beta_n)\geq H_{t,S}(w_n,\beta(w_0))>H_{t,S}(w_0,\beta(w_0))-\epsilon$. Taking the limit and using that $(w,b)\to H_{t,S}(w,b)$ is continuous since its a composition of continuous functions, we conclude that $H_{t,S}(w_0,b_0)\geq H_{t,S}(w_0,\beta(w_0))$ but by definition of the function $\beta$ we also have that $H_{t,S}(w_0,b_0)\leq H_{t,S}(w_0,\beta(w_0))$. This implies $H_{t,S}(w_0,b_0)= H_{t,S}(w_0,\beta(w_0))$. Since the optimal point corresponding to $w_0$ is unique we get $b_0=\beta(w_0)$ and this proves the continuity of $w\to \beta(w)$ at $w_0$.

The continuity of $\beta_{t,S}$ then implies the continuity of $V_{t,S}$. \qed

\bigskip

\begin{lemma}\label{lemma4}
For all $(t,S)\in {\cal T}$, the function  $V_{t,S}$ is concave on its domain $\Re$. 
\end{lemma}
{\bf Proof.}\ The function $w\to (w,\beta(w))$ moves continuously on the grid $G$ by Lemmas \ref{lemma2} and \ref{lemma3} and $w\to\beta(w)$ is linear between two intersections of grid lines. We therefore only need to establish concavity at points $w$ where $(w,\beta(w))$ is on such an intersection, since in all other points, $V_{t,S}$ is the sum of two compositions of concave functions with (fixed) linear functions and therefore concave. 

 Let  $w_0$ be a value which corresponds to a point of intersection: $(w_0,\beta(w_0))\in (C_i^u\cap C_j^d)$ for some $i \in \{ 1, \ldots, N_u \}$ and $j \in \{ 1, \ldots, N_d \}$. In order to establish the concavity of $V_{t,S}$ in $w_0$, we compute the left-hand side and right-hand side derivatives of  $V_{t,s}$ at this point.

If $w$ is larger than $w_0$ but close enough to $w_0$ we must have that $(w,\beta(w))\in C_i^u$ or $(w,\beta(w))\in C_j^d$ which means that $\sfrac{d\beta^+}{dw}(w_0)$ equals $-R/(u-R)$ or $R/(R-d)$ respectively, by \eqref{eq:cu} and \eqref{eq:cd}.
From \eqref{eq:2} we know that
\be
V_{t,S}(w)&=&R^{-1} \left( \  
pV_{t+\Delta t,uS}(w R+\beta(w)(u-R)) \right.\\&& \qquad\quad +\ 
\left. (1-p)V_{t+\Delta t,dS}(wR+\beta(w)(d-R))\ \right),
\ee
and we notice that the first term on the right-hand side is constant for $(w,\beta(w))\in C_i^u$ while  the second term is constant for $(w,\beta(w))\in C_j^d$.
Thus allows us to conclude that in $w=w_0$, the right-hand side derivative $V_{t,S}^{'+}(w_0)$ equals either $R^{-1}(1-p)V^{'+}_{t+\Delta t,dS}(x_j^d)\cdot (R+(d-R)\sfrac{-R}{u-R})=V^{'+}_{t+\Delta t,dS}(x_j^d)\cdot \sfrac{1-p}{1-q}$ in the first case or $R^{-1}pV^{'+}_{t+\Delta t,uS}(x_i^u)\cdot(R+(u-R)\sfrac{R}{R-d})=V^{'+}_{t+\Delta t,uS}(x_i^u)\cdot \sfrac{p}{q}$ in the second case. Since we chose the optimal strategy, it should equal the maximum of the two:
\begin{eqnarray}\label{eq:vplus}
V_{t,S}^{'+}(w_0) &=&\max\ \{ \ \sfrac{1-p}{1-q} V^{'+}_{t+\Delta t,dS}(x_j^d),\  \sfrac{p}{q}V^{'+}_{t+\Delta t,uS}(x_i^u) \ \}.
\end{eqnarray}
To find the left-hand side derivative, we consider values of $w$ that are smaller than $w_0$ but close enough to $w_0$. Again, we must have that $\sfrac{d\beta^-}{dw}(w_0)$ equals $-R/(u-R)$ or $R/(R-d)$ and
by again calculating the value of $V^{'-}_{t,S}$ for these two possibilities and now  choosing the smallest of the two, we find:
\begin{eqnarray}\label{eq:vmin}
V_{t,S}^{'-}(w_0) &=&\min\ \{ \ \sfrac{1-p}{1-q} V^{'-}_{t+\Delta t,dS}(x_j^d),\  \sfrac{p}{q}V^{'-}_{t+\Delta t,uS}(x_i^u) \ \}.
\end{eqnarray}
We must now show that $V^{'+}_{t,S}(w_0)\leq V^{'-}_{t,S}(w_0)$ by comparing the right-hand sides of \eqref{eq:vplus} and \eqref{eq:vmin} to finish the proof.

This can be established by the optimality condition for our choice of $\beta(w_0)$. Since this choice was determined by maximizing the piecewise linear function $b\to H(w_0,b)$ we must have that $\sfrac{\partial^+}{\partial b}H_{t,S}(w_0,\beta(w_0))\leq 0\leq \sfrac{\partial^-}{\partial b}H_{t,S}(w_0,\beta(w_0))$ and at least one of the inequalities must be strict. Since
$$\sfrac{\partial^+}{\partial b}H_{t,S}(w_0,\beta(w_0))=R^{-1}pV^{\prime +}_{t+\Delta t,uS}(x_i^u)(u-R)+R^{-1}(1-p)V^{\prime -}_{t+\Delta t,dS}(x_j^d)(d-R)$$
$$\sfrac{\partial^-}{\partial b}H_{t,S}(w_0,\beta(w_0))=R^{-1}pV^{\prime -}_{t+\Delta t,uS}(x_i^u)(u-R)+R^{-1}(1-p)V^{\prime +}_{t+\Delta t,dS}(x_j^d)(d-R)$$
we have
\be
\sfrac{p}{q}{V^{\prime +}_{t+\Delta t,uS}}(x_i^u) &\leq& \sfrac{1-p}{1-q}{V^{\prime -}_{t+\Delta t,dS}}(x_j^d),\\
\sfrac{1-p}{1-q}{V^{\prime +}_{t+\Delta t,dS}}(x_j^d)&\leq& \sfrac{p}{q}{V^{\prime -}_{t+\Delta t,uS}}(x_i^u) ,
\ee
with at least one of the two inequalities strict. But we also have that
\be
V^{\prime +}_{t+\Delta t,uS}(x_i^u)&<&V^{\prime -}_{t+\Delta t,uS}(x_i^u),\\
V^{\prime +}_{t+\Delta t,dS}(x_j^d)&<&V^{\prime -}_{t+\Delta t,dS}(x_j^d),
\ee
 by concavity of $V_{t+\Delta t,uS}$ and $V_{t+\Delta t,uS}$. Using these four  inequalities to compare the right-hand sides of \eqref{eq:vplus} and \eqref{eq:vmin}, we establish that $V^{'+}_{t,S}(w_0)\leq V^{'-}_{t,S}(w_0)$ holds and this finishes the proof. \qed

\bigskip

\begin{theorem}\label{the:1}
We have $V_{t+\Delta t,uS},V_{t+\Delta t,dS}\in{\cal H}\Rightarrow V_{t,S}\in{\cal H}$.
\end{theorem}

{\bf Proof.}\ We check the properties of Definition \ref{def:1} for $V_{t,S}$: piecewise linearity and differentiability outside a finite set follow from the fact that $(w,\beta(w))\in G$ for all $w$, the fact that $V_{t,S}$ is linear for points in $G$ where $(w,\beta(w))$ is not in the subset of $G$ of intersection points $\cup_i\cup_j (C^u_i\cap C^d_j)$, and the fact that the set of intersections is finite. 
That $V_{t,S}$ is concave was shown in Lemma \ref{lemma3}. The fact that $V_{t,S}$ eventually becomes constant follows from Lemma \ref{lemma2a}, since it is shown there that for $w$ large enough $(w,\beta(w))\in L_{N_u,N_d}^u$. This is because $H_{t,S}$ is constant on $L_{N_u,N_d}^u$, since it is a linear combination of the functions $V_{t+\Delta t,uS}$ and $V_{t+\Delta t,dS}$ which are evaluated in points beyond their last singular value, where they are constant.
 \qed

\bigskip

From the proofs of the Theorem and the Lemmas preceding it, we can now easily derive an algorithm to determine the value functions and optimal investment strategy in every state on the tree.

\medskip

\begin{corollary}
Assume that $V_{t+\Delta t,uS}$ and $V_{t+\Delta t,dS}$ have singular values $x^u_1<x^u_2<...<x^u_{N_u}$ and $x^d_1,x^d_2<...<x^d_{N_d}$ respectively, and let  $$v^u_i=V_{t+\Delta t,uS}(x^u_i),\qquad v^d_i=V_{t+\Delta t,dS}(x^d_i),\qquad
   z^u_i=V^{ '-}_{t+\Delta t,uS}(x^u_i),\quad z^d_i=V^{ '-}_{t+\Delta t,dS}(x^d_i).$$

Then $x_k$, the singular values of $V_{t,S}$ {\bf in reverse order}, its values $v_k=V_{t,S}(x_k)$ in these points, the optimal strategies $\beta_k=\beta_{t,S}(x_k)$ and the left-hand side derivatives $z_k=V_{t,S}^{'-}(x_k)$ satisfy,  for $k\geq 1$,
\begin{equation}
\begin{array}{rclrcl}
x_{k} &=& R^{-1}(qx^u_{i_{k}}+(1-q)x^d_{j_{k}}),&
\beta_{k} &=&(x^u_ {i_k} - x^d_{j_k})/(u-d),\\
z_u &=& \sfrac{p}{q}\, z_{i_k}^u,\qquad\qquad &z_d &=&   \sfrac{1-p}{1-q}\, z_{j_k}^d,\\
z_{k} &=& z_u \wedge z_d,\qquad\qquad &v_{k} &=& v_{k-1} - z_{k-1}(x_{k}-x_{k-1}), \\
i^u_{k+1} &=& i^u_k - {\bf 1}_{\{z_k= z_u \}},\qquad \qquad&
i^d_{k+1} &=& i^d_k - {\bf 1}_{\{z_k=z_d\}}.
\end{array}
\end{equation}
with $i^u_1=N_u$, $i^d_1=N_d$, $v_0=R^{-1}(pv^u_{N_u}+(1-p)v^d_{N_d})$,  and $z_0=x_0=0$. The sequence of singular values in reverse order stops when either $i^u_k=0$ or $i^d_k=0$ for certain $k$. \label{cor:1}
\end{corollary}

{\bf Proof.}\ We know that the points $(x_k,\beta_k)$ must take the form $C^u_{i^u_k}\cap C^d_{i^d_k}$ since they must be on the grid $G$  by Lemma \ref{lemma2a} and the derivative of $V_{t,S}$ should not be continuous in the points $x_k$ so they must be on the intersections of the grid. We also have by Lemma \ref{lemma2a} that the line $L^u_{N_u,N_d}$ is in ${\cal B}$ so the intersection point on the grid with the highest wealth value must be $C^u_{N_u}\cap C^d_{N_d}$, by the continuity of $w\to (w,\beta(w))$. 
 This gives  the first singular values in reverse order, so
 $i^u_1=N_u$ and $i^d_1=N_d$ and $v_1=R^{-1}(pv^u_{N_u}+(1-p)v^d_{N_d})$.
Direct calculation shows that intersection points on the grid are given by $$C^u_{i^u_k}\cap C^d_{i^d_k}=( \ R^{-1}(qx^u_{i^u_{k}}+(1-q)x^d_{i^d_{k}}),\ (x^u_ {i^u_k} - x^d_{i^d_k})/(u-d)\ ),$$ so we are done when we prove that 
$z_{k}= z_u \wedge z_d$. But this follows from \eqref{eq:vmin}.
 \qed

\subsection{Reducing the Number of Singular Values}\label{sect:elim}

The technique for obtaining the value function in every point of the tree gives exact results but it is computationally expensive since at every time step the number of singular values doubles in every node. However, by using a method introduced in [1], we can reduce the number of points involved in the calculations. This generates an approximation for the value function that can be calculated faster, while keeping the approximation errors within explicit bounds that we can specify a priori.

The idea is to remove some singular values at every step in order to simplify the computation of the value function, while controlling the error generated by such an elimination procedure. We fix a given maximal level of the error $\epsilon>0$ and we modify every value functions $V_{t,S}$ immediately after it has been calculated, and hence before it is used in calculations for the subsequent time step, by deleting singular values. We do this in such a way that the new value functions $\tilde{V}_{t,S}$ differ at most $\epsilon$ from $V_{t,S}$ on their whole domain. 

More precisely, to construct an approximation for the function $V_{t,S}$ that has singular values $x_i$, $i=1,...,N$, and corresponding values $v_i=V_{t,S}(x_i)$, we delete singular values in the following way.
We set $i_k=1$ for $k=1$ and starting from $x_{i_k}$, we try to find the largest singular value for which the distance in the interval $[x_{i_k},x_i]$, between the straight line connecting $(x_{i_k},v_{i_k})$ with $(x_i,v_i)$ and the graph of the function $V_{t,S}$ is less than  $\epsilon$: this means that we
define the set
$$ I=\{ i>i_k: \sup_{x\in [x_{i_k},x_i]} |v_{i_k}+(x-x_{i_k})\sfrac{v_i-v_{i_k}}{x_i-x_{i_k}}-V_{t,s}(x)|<\epsilon\}.$$ 
and set
$ i_{k+1}=1+i_k$ if this set is empty and $i_k=\max\{i: i\in I\}$ if it is not empty.

Notice that the distance between $V_{t,S}$ and the straight line through $(x_{i_k},v_{i_k})$ and $(x_i,v_i)$  increases as $i$ increases, by virtue of the concavity of $V_{t,S}$. After we have determined $i_{k+1}$, we delete all the points $x_{i_k+1},...x_{i_{k+1}}$ and we repeat the procedure for the next value of $k$. We continue until the penultimate singular value is reached. We do not modify the function $V_{t,S}$ before the first and after the last singular value.

Using this procedure we obtain a new value function $\tilde{V}_{t,S}$ which is again
in the class $\mathcal{H}$ but with a reduced number of singular values and which differs from $V_{t,S}$ less than $\epsilon$ on its whole domain. Since we have that, for every $w$:
$$|V_{t+\Delta t,dS}(w)-\tilde{V}_{t+\Delta t,dS}(w)|<\epsilon, \qquad |V_{t+\Delta t,uS}(w)-\tilde{V}_{t+\Delta t,uS}(w)|<\epsilon,$$
implies that the new value function in $(t,S)$ will be
\be
\lefteqn{
\tilde V_{t,S}(w)
}&&\\
&=&R^{-1}\sup_{b\in {\Bbb R} }\ \left( \ 
p\tilde{V}_{t+\Delta t,uS}(wR+b(u-R)) \ +\ (1-p)\tilde{V}_{t+\Delta t,dS}(wR+b(d-R))
\ \right)\\
&\leq& R^{-1} \sup_{b\in {\Bbb R} }\ \left( \ 
p[V_{t+\Delta t,uS}(wR+b(u-R))+\epsilon ]
+(1-p)[V_{t+\Delta t,dS}(wR+b(d-R)) +\epsilon ]
\ \right)\\
&=&V_{t,S}(w)+\epsilon. 
\ee
Using a simular argument we find that $\tilde V_{t,S}(w)\geq V_{t,S}(w)-\epsilon$ and
hence we have that $|V_{t,S}(w)-\tilde{V}_{t,S}(w)|<\epsilon$ for all $w$.

The optimal strategy $\beta_{t,S}(w)$ for wealth $w$ that is obtained after replacing $V_{t+\Delta t,uS}$ and $V_{t+\Delta t,dS}$ by $\tilde{V}_{t+\Delta t,uS}$ and $\tilde{V}_{t+\Delta t,dS}$, respectively, can be different from the case where no substitutions have been made, but the value $\tilde{V}_{t,S}(w)$ that it generates differs $\epsilon$ or less from the original value $V_{t,S}(w)$.

\subsection{Extending the number of Risk Factors} 

 We now consider the case where there is another  risk factor, $Y$, which is not tradeable, but which may influence the terminal value of a contingent claim which is deducted from the final wealth.

We therefore now define a multi-nomial tree with nodes that are characterized by $(t,S,Y)$  with $S$ still representing the stock value for a stock that can be traded, and $Y$ a factor which cannot be traded but may be correlated to $S$ or drive the dynamics of $S$ (while still leading to two different return values at every step) such as the stochastic volatility in a Heston Model.

We now have 
\begin{eqnarray}
{\cal T}=\displaystyle\bigcup_{m=0}^n {\cal T}_m,\qquad {\cal T}_m=\bigcup_{k=0}^m\bigcup_{l=0}^m\{(m\Delta t, S_k^m,Y_{l}^m)\}. \label{eq:newcalT}
\end{eqnarray}
The dynamics for the risky asset $S$ remain as defined earlier, but the values of $Y_l^m$ may have a different structure.

At every time step the nodes that can be reached from  $(t,S,Y)$ are 
 $(t+\Delta t,uS,\tu Y)$, $(t+\Delta t,uS,\td Y)$, $(t+\Delta t,dS,\tu Y)$ and $(t+\Delta t,dS,\td Y)$, with probabilities $p_{uu}$, $p_{ud}$, $p_{du}$ and $p_{dd}$ respectively\footnote{As before we suppress some notation; for example, $\td$ is short for $\tilde{d}(t,S,Y)$ and $p_{uu}$ an abbreviation of $p_{uu}(t,S,Y)$ etcetera. Notice that $p_{uu}$ is itself shorthand notation for the probability that both $S$ and $Y$ attain the "upper"  values of their two possibilities in the next time step.}. We assume that $Y$ is a riskfactor that cannot be traded, so we still only allow investment in stock $S$ and cash.

 In this section, our state is $X=(S_t,Y_t,W_t)$ and we will write $\tilde{V}_{t,X_t}=V_{t,S_t,Y_t}(W_t)$ and $\beta_{t,S_t,Y_t}(W_t)$ for the smallest value which makes the strategy $\phi$ defined by\\ $\phi_t(X_t)=\beta_{t,S_t,Y_t}(W_t)$ optimal. 
This implies that
\be
V_{t,S,Y}(w)&=&\max_{b\in \Re} H_{t,S,Y}(w,b),\\ B_{t,S,Y}(w)&=& \underset{b \in \Re}{\arg\max} \ H_{t,S,Y}(w,b), \qquad  \beta_{t,S,Y}(w)=\min \{b:b\in B_{t,S,Y}(w)\},
\ee
with
\begin{equation}\label{eq:2a}
\begin{array}{rcll}
\lefteqn{H_{t,S,Y}(w,b)=} && \\
&&R^{-1}( 
&p_{uu}V_{t+\Delta t,uS,\tu Y}(wR+b(u-R)) \ +\ 
p_{ud}V_{t+\Delta t,uS,\td Y}(wR+b(u-R))+ \\
&&&p_{du}V_{t+\Delta t,dS,\tu Y}(wR+b(d-R)) \ +\ 
p_{dd}V_{t+\Delta t,dS,\td Y}(wR+b(d-R))
\quad ).
\end{array}
\end{equation}

We define $p_u=p_{uu}+p_{ud}$ and $p_d=p_{du}+p_{dd}=1-p_u$ and write
$$
H_{t,S,Y}= R^{-1}\ (  \ 
p_u V_{t+\Delta t,uS}(wR+b(u-R)) \ +\ 
p_d V_{t+\Delta t,dS}(wR+b(d-R)) \ ).
$$
for functions 
\be
 V_{t+\Delta t,uS}(w) &=&   \sfrac{p_{uu}}{p_u} V_{t+\Delta t,uS,\tu Y}(w) \   + \   \sfrac{p_{ud}}{p_u}V_{t+\Delta t,uS,\td Y}(w),\\
 V_{t+\Delta t,dS}(w) &=&   \sfrac{p_{du}}{p_d}V_{t+\Delta t,dS,\tu Y}(w) \   + \   \sfrac{p_{dd}}{p_d}V_{t+\Delta t,dS,\td Y}(w).
\ee
These two functions are in ${\cal H}$ and they have singular values $x^u_i$ and $x^d_i$ which can be found by combining the singular values of $V_{t+\Delta t,uS,\tu Y}$  and $V_{t+\Delta t,uS,\td Y}$ and by combining those of  $V_{t+\Delta t,dS,\tu Y}$  and $V_{t+\Delta t,dS,\td Y}$  respectively. The corresponding values $v^u_i$ and $v^d _i$ can easily be determined by summing the corresponding values of the constituting functions. We are then back in the situation of Theorem \ref{the:1} since we can then write
\be
H_{t,S,Y}(w,b)&=&R^{-1}\ (\  p_{u} V_{t+\Delta t,uS}(wR+b(u-R))\\&&\qquad \ +\ 
(1-p_{u}) V_{t+\Delta t,dS}(wR+b(d-R)) \ ).
\ee
for functions $ V_{t+\Delta t,uS}$ and $ V_{t+\Delta t,dS}$ which are both in ${\cal H}$. This means that we can use the algorithm described in Corollary \ref{cor:1} to calculate the value function in each point of the tree.

 This construction shows that we can treat the case where the stock dynamics depends on an untradeable factor. We now show how this can be exploited to treat an optimal portfolio problem which involves the stochastic volatility model for equity prices that was proposed  by Heston \cite{Heston}. In that model,
the squared volatility process $\tilde{Y}$  and log stock price process $\ln\tilde{S}$ are assumed to satisfy the stochastic differential equations
\begin{eqnarray}
d\tilde Y_t &=& \kappa(\theta-\tilde Y_t)dt \ + \ \omega \sqrt{\tilde Y_t}dW_t^1,\\
d(\ln\tilde S_t) &=& (\mu-\sfrac12 \tilde Y_t) dt \ + \ \sqrt{\tilde Y_t}dW_t^2, \label{eq:HestondS}
\end{eqnarray}
for given $(\tilde Y_0,\ln \tilde S_0)=(\sigma^2,\ln s_0)$,
where $\{ (W_t^1,W_t^2),\ t\in [0,T]\}$ are correlated standard Brownian Motion processes with correlation coefficient $\rho$ , and $\mu$, $\kappa$, $\omega$, $\theta$ and $r$ are strictly positive constants.

We now define the stochastic processes $(Y_m,\ln S_m)$, at times $m=0...n-1$ as the discrete counterpart to the continuous time process $(\tilde{Y}_t,\ln \tilde{S}_t)$:
\begin{eqnarray}\label{eq:pp}
Y_{m+1} &=& Y_m+\kappa(\theta-(Y_m)^+)\Delta t + \eta_{m+1}^{Y}\omega\sqrt{(Y_m)^+\Delta t},\\
\ln S_{m+1} &=& \ln S_m+(\mu-\sfrac12 Y_m)\Delta t + \eta_{m+1}^{S}\sqrt{(Y_m)^+\Delta t}, \nonumber
\end{eqnarray}
where the variables $(\eta^{S}_m,\eta^{Y}_m)$ are i.i.d. distributed in $m$, with\footnote{Notice that we have taken the positive part of the variance process $(Y_m)^+=\max\{0,Y_m\}$ to ensure positivity of this process. This choice corresponds to the full truncation scheme of Lord et al. in \cite{LordKoekkoekVanDijk}.}
\begin{eqnarray*}
p_{1,1}:={\Bbb P}(\eta_{m}^{S}=+1,\ \eta_{m}^{Y}=+1)&\ =\ &p_{-1,-1}:={\Bbb P}(\eta_{m}^{S}=-1,\ \eta_{m}^{Y}=-1) \ = \ \sfrac14(1+\rho)\\
p_{-1,1}:={\Bbb P}(\eta_{m}^{S}=-1,\ \eta_{m}^{Y}=+1)&\ =\ &p_{1,-1}:={\Bbb P}(\eta_{m}^{S}=+1,\ \eta_{m}^{Y}=-1) \ = \ \sfrac14(1-\rho).
\end{eqnarray*}

This generates a tree which is not recombining, and we therefore modify it as follows.
Let $S_m^{{\rm max}}=\max\{s: {\Bbb P}(S_m=s)>0 \}$ and define $S_m^{{\rm min}}$, $Y_m^{{\rm max}}$ and $Y_m^{{\rm min}}$ analogously. We take 
$\Delta S_m=(S_m^{{\rm max}}-S_m^{{\rm min}})/m_z$ and 
$\Delta Y_m=(Y_m^{{\rm max}}-Y_m^{{\rm min}})/m_v$
for certain $m_v,m_z\in {\Bbb N}^+$ which describe how fine the mesh is that we will take, and then define the set of tree nodes $\cal T$ in \eqref{eq:newcalT} using
$S_k^m=S_m^{{\rm min}}+k\Delta S_m$ and 
$Y_l^m=Y_m^{{\rm min}}+l\Delta Y_m$. 

A node $(t,S,Y)$ on the tree, which represents the state of our dynamic process,  must have the form $(t,S,Y)=(m\Delta t,S_k^m,Y_l^m)$ for some $m$, $k$ and $l$. From this state, transitions are possible   to four possible new states of the form $(t+\Delta t,\, S R^{S,Y}_{ \eta_{m+1}^{S}}, Y\tilde{R}^{S,Y}_{ \eta_{m+1}^{Y}})$ for $g\in{\cal U}$ where
\be
 R^{S,Y}_{\eta_{m+1}^{S}} &=& \exp(\ (\mu-\sfrac12 Y)\Delta t + \eta_{m+1}^{S}\sqrt{Y^+\Delta t}\ ),\\
\tilde{R}^{S,Y}_{ \eta_{m+1}^{Y}} &=&  (\ Y+\kappa(\theta-Y^+)\Delta t + \eta_{m+1}^{Y}\omega\sqrt{Y^+\Delta t}\ )/Y.
\ee

This will in general not give a new state of the form $((m+1)\Delta t,\ S_{k^*}^{m+1},\ Y_{l^*}^{m+1})$ for certain $k^*$ and $l^*$, because the tree is not recombining. But in \cite{VellekoopNieuwenhuisHestonTree} it is shown that weak convergence of the process on the tree to its counterpart in continuous time will still be guaranteed if we use, in each of these four points, linear interpolation based on the four points on the grid with the smallest distance to the intended location. We follow this approach here as well and therefore determine $k^o$ and $l^o$ such that 
$$S_{k^S(\eta)}^{m+1}\leq S R^{S,Y}_{\eta} \leq S_{k^S(\eta)+1}^{m+1},\qquad  Y_{k^Y(\eta)}^{m+1}\leq Y\tilde{R}^{S,Y}_{\eta}\leq  Y_{k^Y(\eta)+1}^{m+1},$$ 
for $\eta\in\{-1,1\}$ and use the linear combination
\begin{eqnarray}\label{eq:approxeta}
V_{t+\Delta t, S R^{S,Y}_{\eta^{S}} , Y\tilde{R}^{S,Y}_{\eta^{Y}} }&=&
\sum_{i=0}^1\sum_{j=0}^1 \widetilde{p}_{ij}V_{(m+1)\Delta t, S_{k^S(\eta^S)+i}^{m+1} , Y_{k^Y(\eta^Y)+j}^{m+1}}
\end{eqnarray}
with linear interpolation weights
\be
\widetilde{p}_{11}&=&
\frac{S        R^{S,Y}_{\eta^{S}}-S_{k^S(\eta^S)}^{m+1}}{\Delta S_{m+1}}\cdot
\frac{Y\tilde{R}^{S,Y}_{\eta^{Y}}-Y_{k^Y(\eta^Y)}^{m+1}}{\Delta Y_{m+1}},\qquad\\
\widetilde{p}_{00} &=& 
\frac{S_{k^S(\eta^S)+1}^{m+1}-S        R^{S,Y}_{\eta^{S}}}{\Delta S_{m+1}}\cdot
\frac{Y_{k^Y(\eta^Y)+1}^{m+1}-Y\tilde{R}^{S,Y}_{\eta^{Y}}}{\Delta Y_{m+1}},\\
\widetilde{p}_{01}&=&
\frac{S_{k^S(\eta^S)+1}^{m+1}-S        R^{S,Y}_{\eta^{S}}}{\Delta S_{m+1}}\cdot
\frac{Y\tilde{R}^{S,Y}_{\eta^{Y}}-Y_{k^Y(\eta^Y)}^{m+1}}{\Delta Y_{m+1}},\qquad\\
\widetilde{p}_{10} &=&
\frac{S        R^{S,Y}_{\eta^{S}}-S_{k^S(\eta^S)}^{m+1}}{\Delta S_{m+1}}\cdot
\frac{Y_{k^Y(\eta^Y)+1}^{m+1}-Y\tilde{R}^{S,Y}_{\eta^{Y}}}{\Delta Y_{m+1}}.
\ee
The weights can be interpreted as new probabilities , since the four different transitions from the current state $(t,S,Y)$ are each divided over four future states $(t+\delta t,\tilde{S},\tilde{Y})$ each, to create sixteen transitions in total. On our tree, we implement this by simply taking a linear combination of four value functions in class ${\cal H}$, since this results in a function which is again in class $\cal H$, and this function can be determined very efficiently using the associated set of singular values. Figure \ref{fig:hestongrid} illustrates the construction: four transitions to points which may not be on the grid are replaced by sixteen transitions which are on the grid and since the weights are positive and sum to one, these can be interpreted as a new set of probabilities on the tree.

\begin{figure}[t!]
\begin{center}
 \includegraphics[width=10cm]{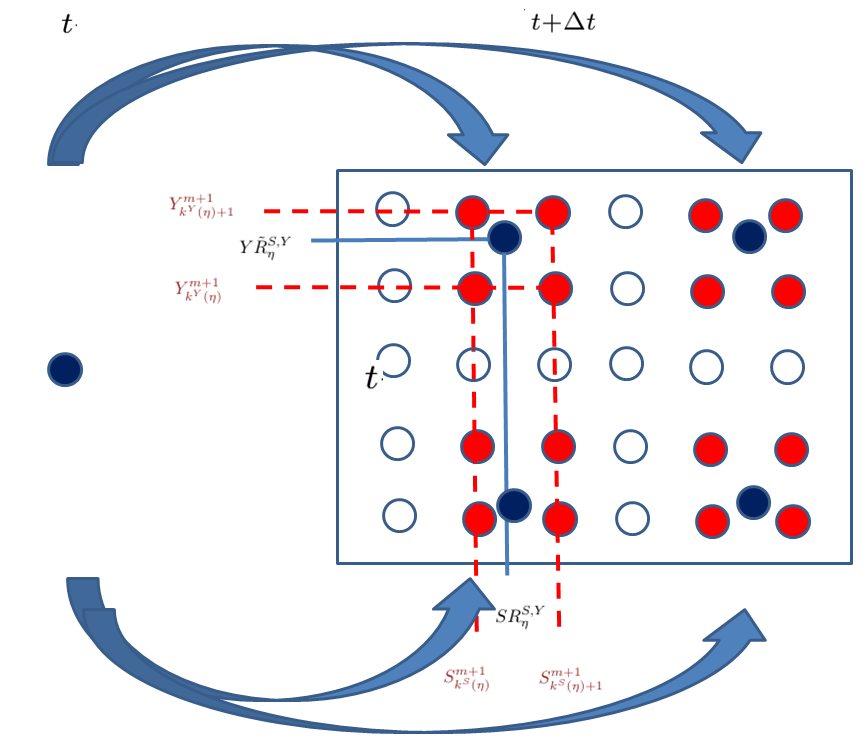}
 \caption{Each of the four succesor value functions (in black) is based on four nearby valuation functions (in red).}
\label{fig:hestongrid}
 \end{center}
 \end{figure}

We now write \eqref{eq:2a}, using \eqref{eq:approxeta}, as
\begin{eqnarray*}
\lefteqn{H_{t,S,Y}(w,b)}&&\\ &=& R^{-1}
\sum_{\eta^S\in\{-1,1\}}\sum_{\eta^V\in\{-1,1\}} p_{\eta^S,\eta^V}
V_{t+\Delta t,SR^{S,Y}_{\eta^S},S\tilde{R}^{S,Y}_{\eta^Y}}(wR+( R^{S,Y}_{\eta^{S}}-R)b)\\
&\approx& 
R^{-1}\sum_{\eta^S\in\{-1,1\}}\sum_{\eta^V\in\{-1,1\}} p_{\eta^S,\eta^V}
\sum_{i=0}^1\sum_{j=0}^1 \widetilde{p}_{ij}V_{t+\Delta t, S_{k^S(\eta^S)+i}^{m+1} , Y_{k^Y(\eta^Y)+j}^{m+1}}(wR+( R^{S,Y}_{\eta^{S}}-R)b)\\
&=& R^{-1}\sum_{\eta^S\in\{-1,1\}} \bar{p}_{\eta^S}\bar{V}_{t+\Delta t, \eta^S}(wR+( R^{S,Y}_{\eta^{S}}-R)b)
\end{eqnarray*}
where
\begin{eqnarray*}
\bar{V}_{t+\Delta t, \eta^S} &=&
\sum_{\eta^V\in\{-1,1\}}\sum_{i=0}^1\sum_{j=0}^1  p_{\eta^S,\eta^V}
\widetilde{p}_{ij}V_{t+\Delta t, S_{k^S(\eta^S)+i}^{m+1} , Y_{k^Y(\eta^Y)+j}^{m+1}}/\bar{p}_{\eta^S}\\
\bar{p}_{\eta^S} &=& \sum_{\eta^V\in\{-1,1\}}\sum_{i=0}^1\sum_{j=0}^1  p_{\eta^S,\eta^V}
\widetilde{p}_{ij}
\end{eqnarray*}
The two functions $\bar{V}_{t+\Delta t, 1}$ and $\bar{V}_{t+\Delta t, -1}$ are in ${\cal H}$ and they have singular values  which can be found by combining the singular values of the functions $$V_{t+\Delta t, S_{k^S(\eta^S)+i}^{m+1} , Y_{k^Y(\eta^Y)+j}^{m+1}}$$ of which they are a linear combination.  We are then back in the situation of Theorem \ref{the:1} and the algorithm in Corollary \ref{cor:1} can be applied.

\begin{figure}[t!]
\begin{center}
 \includegraphics[width=16cm]{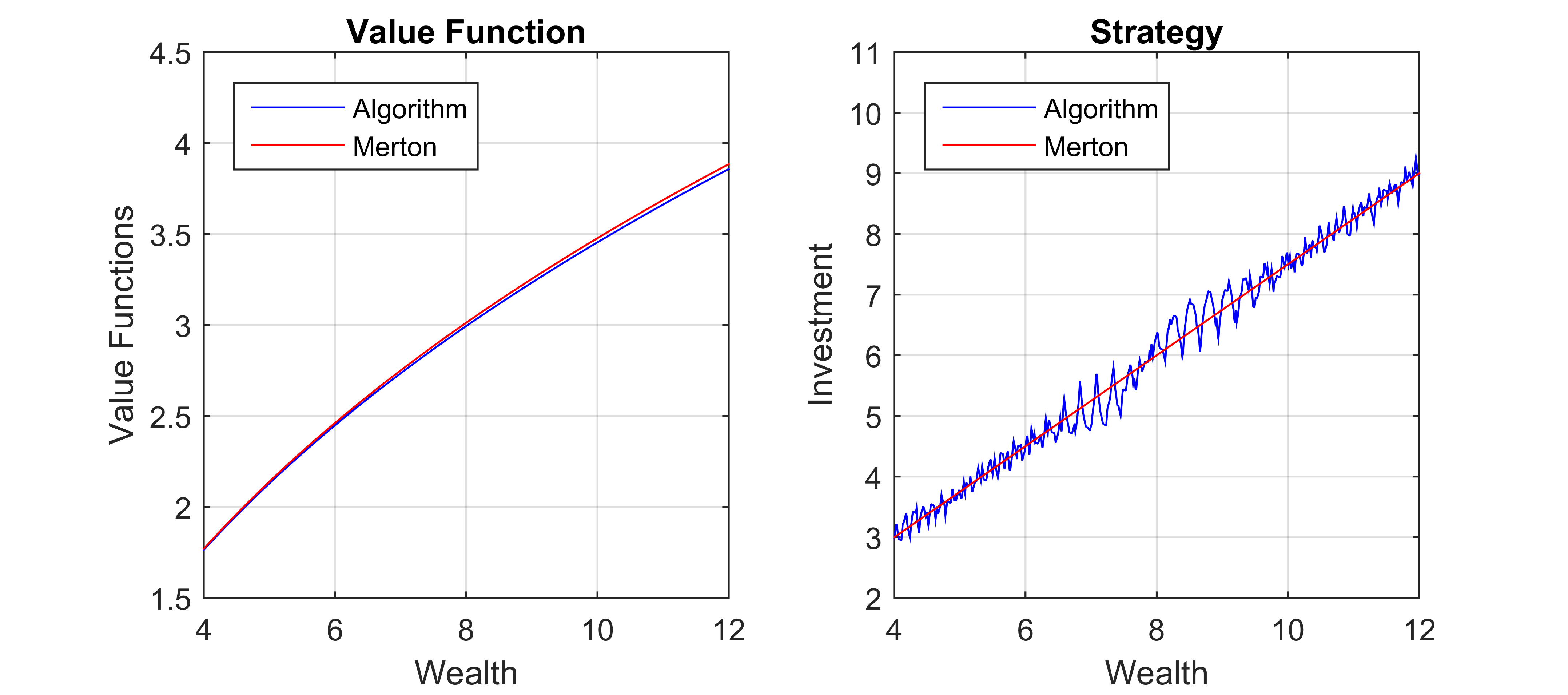}
 \caption{Power Utility Case.}
\label{fig:MertonPower}
 \end{center}
 \end{figure}

\section{Numerical Examples}\label{sec:numericalexamples}

We now apply our method to a set of different investment and indifference pricing problems in both complete and incomplete markets. 

\subsection{Optimal Portfolio Choice}

 In our first numerical example we calculate the optimal strategy and optimal value function for an investor in a Black-Scholes economy with bonds, which earn a constant rate of return $r$ per unit of time, and stocks with a constant mean rate of return $\mu$ and volatility $\sigma$. We consider an investment horizon of $T=1$ (year) and use the standard binomial tree with time steps $\Delta t=T/n$, so $u=1/d=\exp(\sqrt\sigma\Delta t)$, $R=\exp(r\Delta t)$ and $q=(u-R)/(u-d)$ are all constant over time. We take $r=1\%$,  $\sigma=10\%$ and $\mu=1.5\%$, unless otherwise specified.

\subsubsection{Constant Relative Risk Aversion Case}

We first consider the utility function 
\begin{equation}
U(x)=(x^{1-\gamma}-1)/(1-\gamma), \label{eq:powerutility}
\end{equation}
 with constant relative risk aversion (CRRA), for $\gamma=\sfrac23$.

If we take small time steps, the value function in our discrete time setup around the initial wealth value $w_0$ and initial stock value $S_0=5$, i.e. $V^n_{0,S_0}(w)$, should be close to  to the continuous time limit value function derived by Merton \cite{MertonOptimalConsumption}. That value function, and the corresponding optimal investment strategy, are
$$
V^{\rm cont}_{0,S_0}(w) = e^{\xi  T} U(w),\quad
 \beta^{\rm cont}_{0,S_0}(w)=\sfrac{\mu-r}{\gamma\sigma^2}w,\qquad
 \xi = (1-\gamma)(r+\sfrac{(\mu-r)^2}{2\gamma\sigma^2}).
$$
We represent the utility function for terminal wealth using an approximating function which is in class ${\cal H}$, by defining $N_U$ equidistant singular points $(x^U_i)_{i=1...N_U}$ on the interval $[w_{\rm min},w_{\rm max}]$.  We used $N_U=50$ initial singular points, $n=20$ time steps.

Results are shown in Figure \ref{fig:MertonPower}. The red lines correspond to the optimal strategy in the continuous time limit, as derived by Merton, and the blue lines show the results of our algorithm. We note that the value functions are very close, even though the optimal strategies show a rather different behaviour.

\subsubsection{Constant Absolute Risk Aversion Case}
In a similar way, we analyze exponential preferences, which correspond to constant absolute, instead of relative, risk aversion. This means that  $U(x)=-\exp(-\gamma x)$ and
$$
V^{\rm cont}_{0,S_0}(w) = e^{\xi  T} U(we^{rT}),\quad
 \beta^{\rm cont}_{0,S_0}(w)=\sfrac{\mu-r}{e^{rT}\gamma\sigma^2},\qquad
 \xi = -\sfrac{(\mu-r)^2}{2\sigma^2}.
$$
We keep the other parameters the same as in the previous subsection.

Results for this case are shown in Figure \ref{fig:MertonExponential}.

\begin{figure}[t!]
\begin{center}
 \includegraphics[width=16cm]{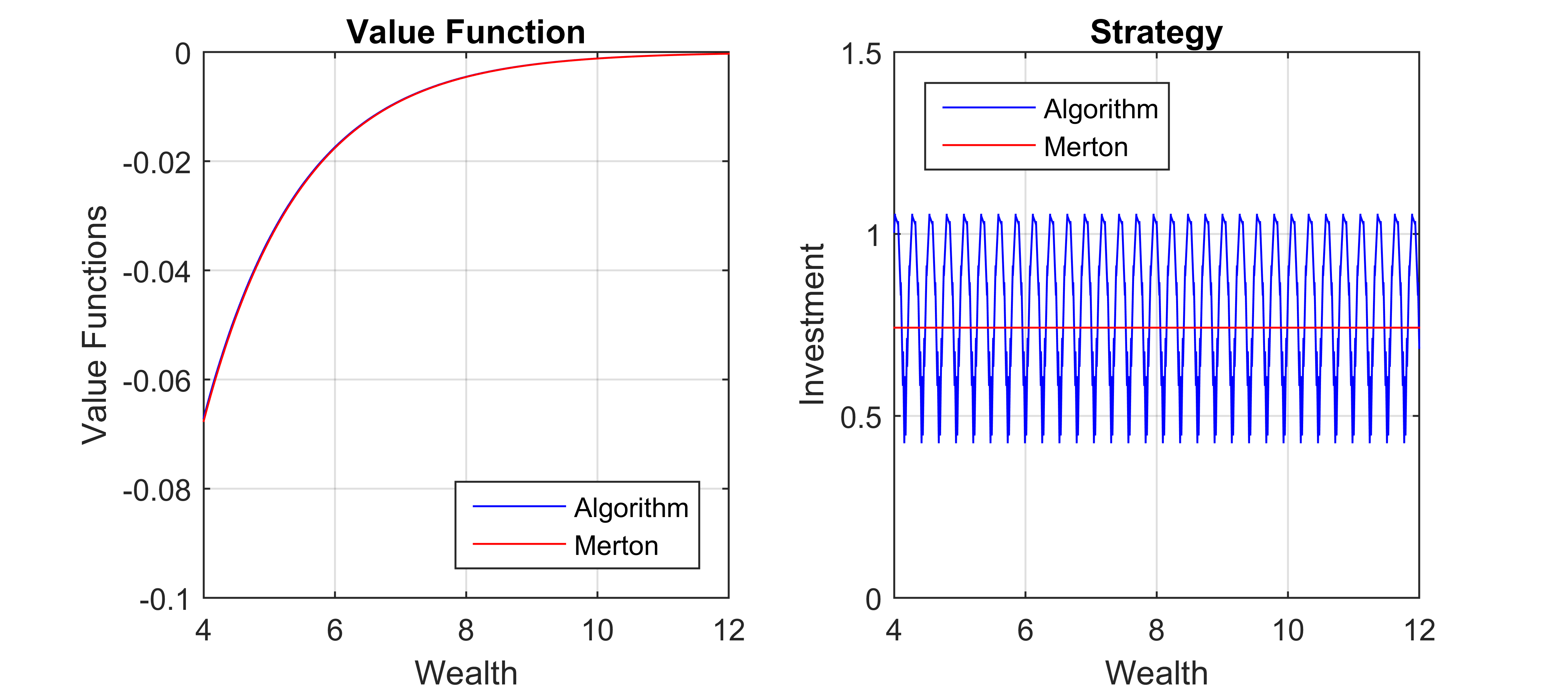}
 \caption{Exponential Utility Case.}
\label{fig:MertonExponential}
 \end{center}
 \end{figure}

\subsection{Indifference Pricing in a Complete Market}

 We can use either the CARA or the CRRA utility function to check the price of vanilla derivatives in the complete market that is generated by our stock price tree and the riskfree asset. 
As mentioned in the introduction, the utility indifference price $\pi$ at time zero of a European option with payoff $\Psi(S_T)$ at a date $T>0$ is the solution to the equation $V_{0,S_0}(w)=V^*_{0,S_0}(w+\pi)$ where $V$ and ${V^c}$ are value functions with $V_{T,S}(w)=U(w)$ and $V^*_{T,S}(w)=U(w-\Psi(S))$ respectively. This means that we assign the same value today to an amount of wealth $w$ without a derivative as to a position where we know that we have to pay the payoff $\Psi(S_T)$ at the maturity date $T$, but get compensated for this by receiving the price $\pi$ to add to our current wealth $w$.

\begin{figure}[t!]
\begin{center}
 \includegraphics[width=16cm]{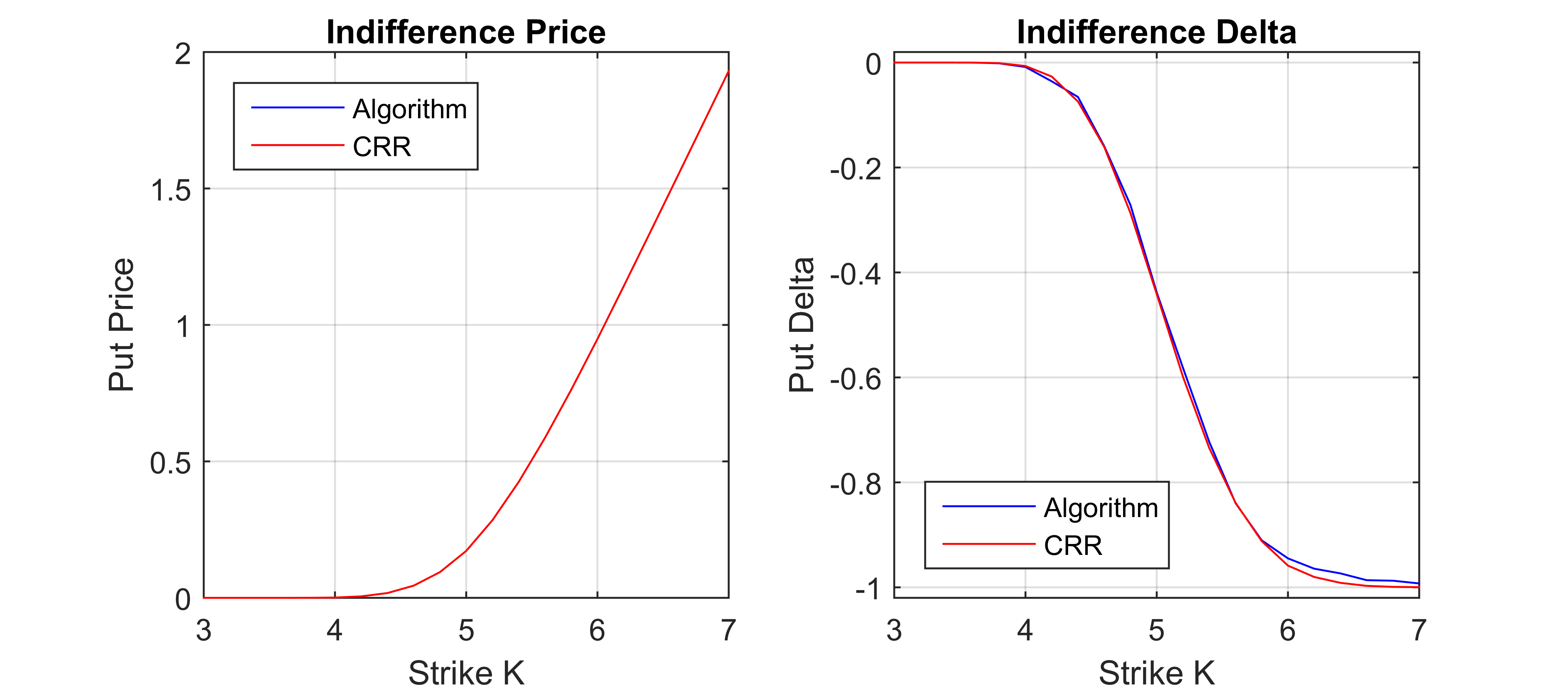}
 \caption{Exponential Indifference Prices and Deltas for Put Options.}
\label{fig:CARAPut}
 \end{center}
 \end{figure}

 In a complete market such as the Cox-Ross-Rubinstein (CRR) model we consider here, we can create a separate portfolio with stocks and bonds which perfectly replicates the payoff and which has an initial price that equals the CRR value of the option. This means that $\pi$ must equal this value, since there is a perfect separation between stock and bonds investments that we use to pay off the option and stocks an bonds that we use to create a portfolio which optimizes the utility function of the rest of the wealth $w$.  Showing that the utility indifference price equals the CRR price thus requires that our algorithm generates the nonlinear optimal investment in stocks that will now be needed, which is known as $\Delta$.   The value of $\pi$ can be calculated from $V_{0,S_0}(w)=V^*_{0,S_0}(w+\pi)$ as $$\pi=(V^*_{0,S_0})^{-1}(V_{0,S_0}(w))-w,$$ which, in our complete market, should be the same for all values of $w$. The corresponding initial $\Delta$, i.e. the number of stocks that we will invest in to replicate the payoff of a single option, can be found by 
$$\Delta = \frac{ \beta^*_{0,S_0}(w)-\beta_{0,S_0}(w)}{S_0}$$ and should also not depend on $w$.

 As an example, we take put options with maturity $T=1$ and strike prices between $K=3$ and $K=7$, while the current stock price equals $S_0=5$. The results in Figure \ref{fig:CARAPut} for the Exponential Utility Indifference Prices and Deltas show excellent agreement with the corresponding CRR values. We found equally good agreement for other utility functions.

\subsection{Indifference Pricing in an Incomplete Market}

\begin{figure}[t!]
\begin{center}
 \includegraphics[width=16cm]{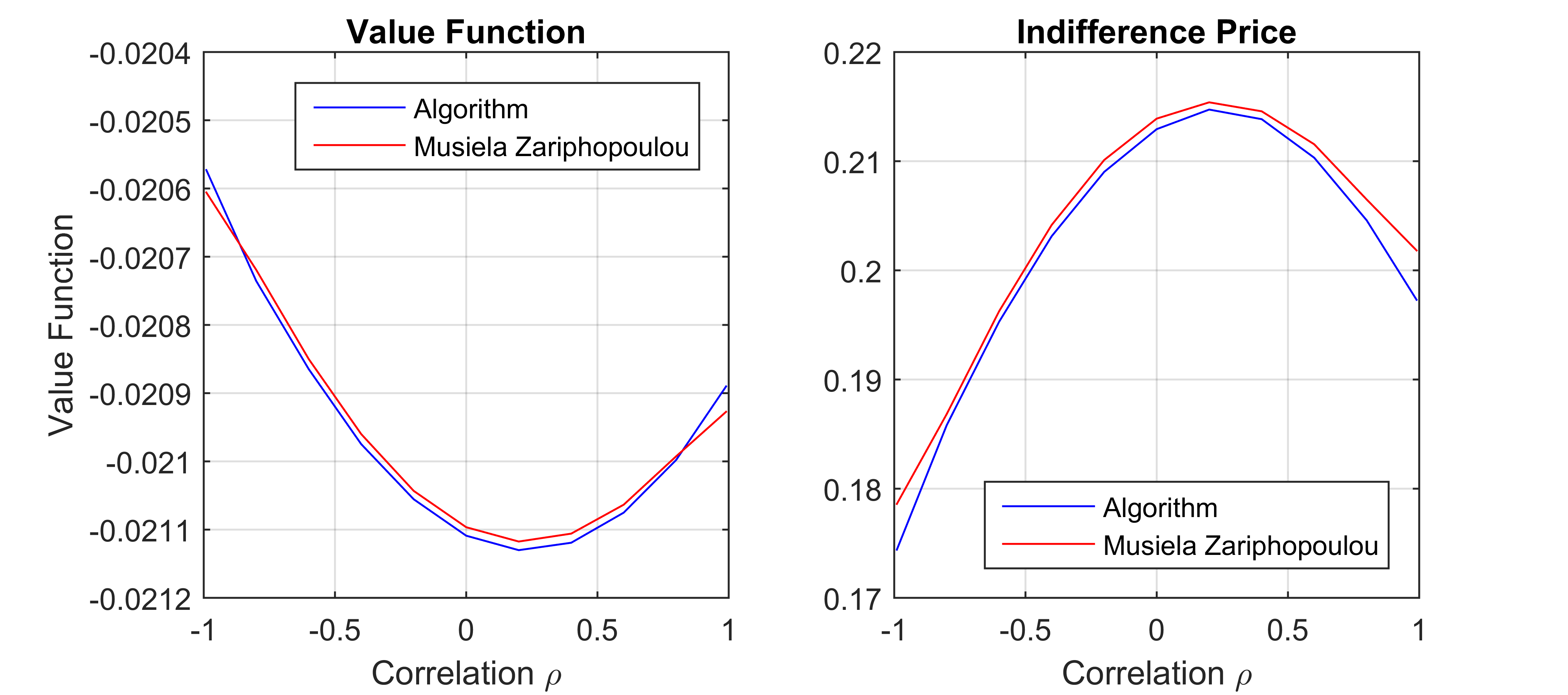}
 \caption{CARA Value Functions and Indifference Prices for Put Options.}
\label{fig:MuZaPut}
 \end{center}
 \end{figure}

 We now turn to an example where the market is incomplete. Since this means that no perfect riskless replication is possible, there is no universal price for derivatives: the utility indifference price that any agent would agree to pay for a certain future payoff will now depend on his or her risk preferences.
 
We use the economy introduced by Zariphopoulou and Musiela model  in \cite{MusielaZariphopoulou2004}, with a tradeable stock and another price process $Y$ which is correlated but untradeable:
\be
dS_t&=&\mu_S S_t dt + \sigma_S S_t dW_t^S, \\
dY_t&=&\mu_Y Y_t dt + \sigma_Y Y_t dW_t^Y, 
\ee
with $d\langle W^S,W^Y\rangle_t=\rho dt$.
They derive an explicit expression for the utility indifference price $\pi$ at time zero for a payoff $G(Y_T)$ at a future time $T>0$ under constant absolute risk aversion preferences $U(x)=-e^{-\gamma x}$ and zero interest rate. This price $\pi$ equals
\begin{equation}
\pi \ = \ \frac{\ln {\Bbb E}[\, e^{\gamma(1-\rho^2)g(Y_T)}\frac{d{\Bbb Q}}{d{\Bbb P}}\, ]}{\gamma(1-\rho^2)},\qquad \frac{d{\Bbb Q}}{d{\Bbb P}}=e^{-\frac{\mu_S}{\sigma_S} W_T^S-\sfrac{\mu_S^2}{2\sigma_S^2}T}.\label{eq:muzapi}
\end{equation}

 In our discrete time version of this model, we used the same parameters for the economy as before, but  zero interest rates, and determine the  price of a times a put option with strike $K=S_0=5$. We find for  $N_u=50$ singular points and $n=20$ timesteps the results in figure \ref{fig:MuZaPut}, which are compared to the theoretical value which is found using Monte Carlo simulations based on (\ref{eq:muzapi}).

\subsection{Optimal Investment under Stochastic Volatility}

\begin{figure}[t]
\begin{center}
 \includegraphics[width=\textwidth]{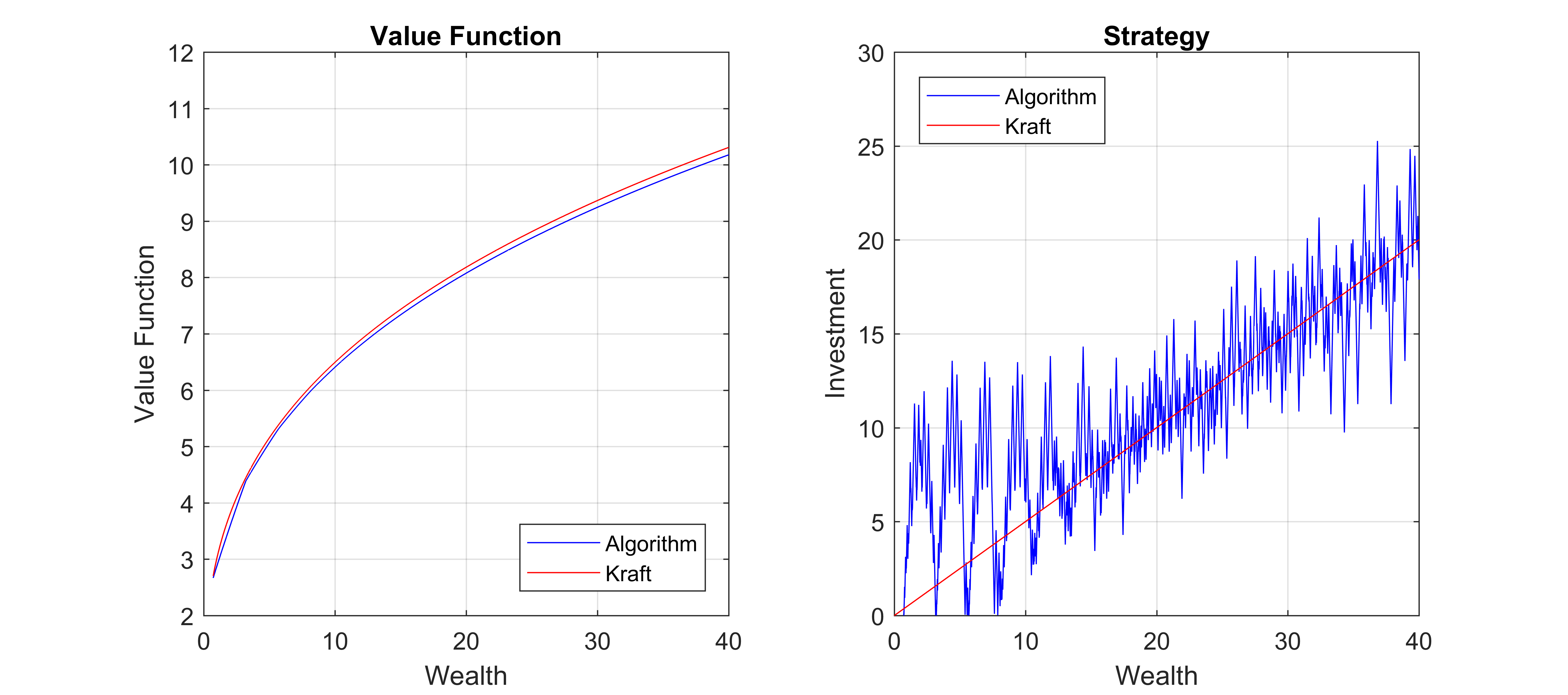}
 \caption{Optimal Strategy derived for (discretized) Heston dynamics.}
\label{fig:Kraft}
 \end{center}
 \end{figure}

 In a paper by Kraft \cite{Kraft} explicit forms are given for the optimal investment strategy in Heston's stochastic volatility model. When $\mu-r= \lambda \tilde{Y}_t$ in \eqref{eq:HestondS} for a given $\lambda>0$ the optimal investment strategy for power utility \eqref{eq:powerutility} equals
$$
b(w)\ = \ w\gamma^{-1}(\lambda + \gamma^{-1}(1-\gamma)\rho\sigma\lambda^2)\, 
\frac{e^{a(T-t)}-1}{e^{a(T-t)}(k+a)+a-k}$$
with $$
k = \kappa - (1-\gamma^{-1})\rho\lambda\sigma,\quad
c = \sfrac{\gamma}{\gamma+\rho^2(1-\gamma)},\quad
B = -\sfrac{\lambda^2(1-\gamma)}{2c\gamma},\quad
a = \sqrt{k^2+2B\sigma^2}.
$$
For the parameter values $r=10\%$, $\sigma=25\%$, maturity $T=0.25$, correlation $\rho=0.10$, volatility of volatility $\omega=39\%$, mean reversion $\kappa=1.15$ and level $\theta=16\%$ we find the figures below for the optimal strategy at different points in time. 

We used risk aversion parameter $\gamma=\sfrac23$ and market price of volatility risk parameter $\lambda=\sfrac{1}{3}$. The graphs in Figure \ref{fig:Kraft} were calculated with $50$ time steps, $125$ gridpoints in the (log) stock price direction and $50$ in the (squared) volatility direction. We used $20$ singular points to start with. We show the results halfway the time to maturity i.e. for $t=T/2$ in the middle of the grid. Other points for $(t,S,Y)$ gave equally good results.

\subsection{An Example of Nonlinear Optimal Policies}

\begin{figure}[t!]
\begin{center}
 \includegraphics[width=16cm]{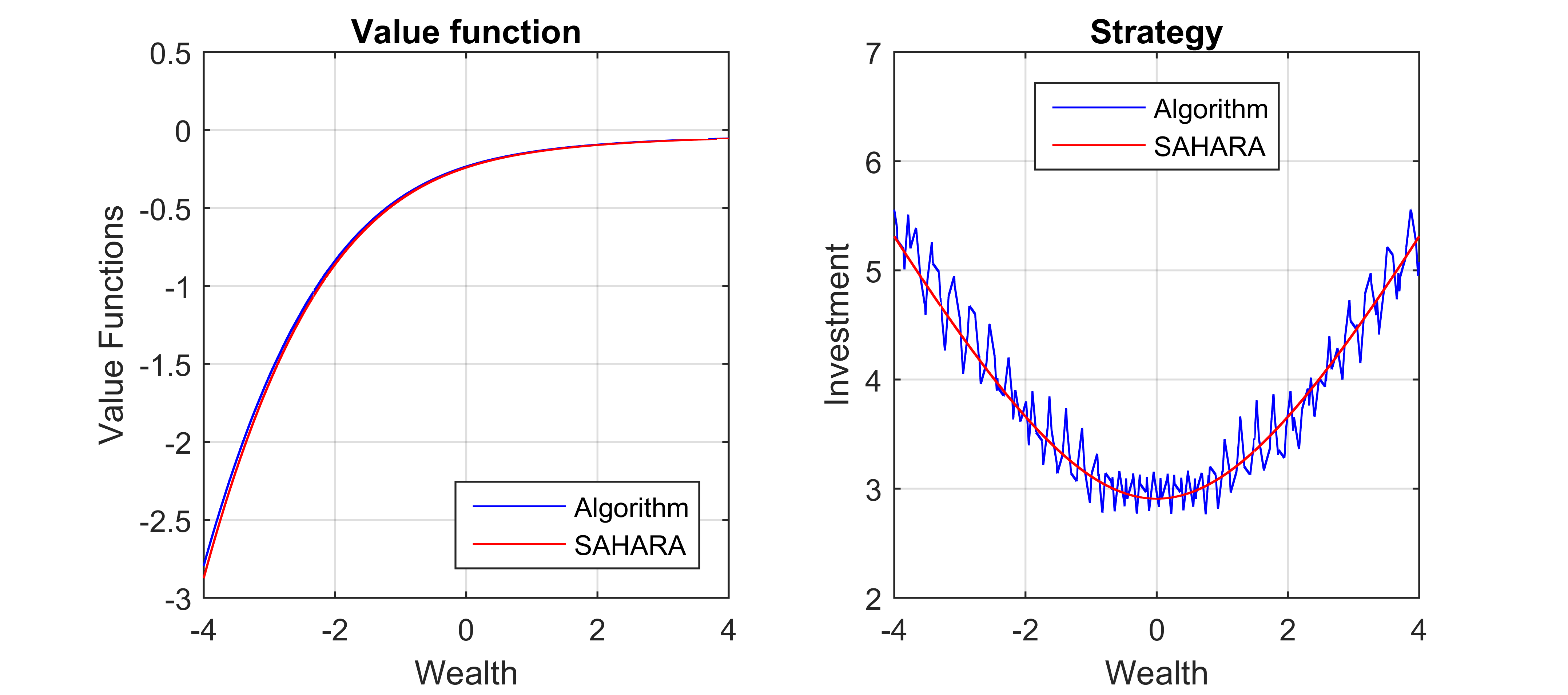}
 \caption{SAHARA case.}
\label{fig:SAHARA}
 \end{center}
 \end{figure}

 In the examples that we have treated so far, closed-form solutions were available for the optimal investment policies. These always took the shape of a linear function of current wealth. We now show that our algorithm can also accurately represent a more complicated strategy, which is optimal for the class of Symmetrically Adjusted Hyperbolic Absolute Risk Aversion preferences (SAHARA). Such preferences are characterized by the utility function
$$
U_{\alpha,\beta}(w)\ = \ 
\frac{1}{1-\alpha^2}\left(w+\,\sqrt{\beta^2+w^2}\right)^{-\alpha}
\left(w+\alpha \,\sqrt{\beta^2+w^2}\right)
$$
where $\alpha>0$ is a risk aversion parameter unequal to one\footnote{Another member of this family of utility functions can be found by taking the limit $\alpha\to 1$ but we will not consider that case here.} and $\beta$ a scaling parameter. 
Under the Black-Scholes dynamics introduced earlier, the optimal strategy turns out to be \cite{SAHARApaper}
$$
\beta_{t,S_t}^{\rm cont}(w) = \frac{(\mu-r) }{\alpha\sigma^2}\sqrt{ w^2 + b(t)^2}, \qquad
b(t) =\beta e^{-(r-\sfrac12((\mu-r)/(\alpha\sigma))^2)(T-t)}.
$$
The corresponding value function satisfies $V_{0,S_0}(w)=U_{\alpha,b(t)}(w)$ so it has the same functional form as the terminal value but with a different, time-dependent scaling parameter.

 Figure \ref{fig:SAHARA} shows the result of a calculation with 
 75 singular points and 20 time steps for $\alpha=2$ and $\beta=2.66$. The Black-Scholes economy parameters were left unchanged from earlier cases.
We find again excellent agreement, both for the value function and the optimal strategy, even though the latter is not linear for this specification.

\section{Conclusions and Further Research}

 We have shown how discrete time optimal investment and indifference pricing problems for asset prices on trees can be solved exactly if one assumed that the utility function which describes the investor's preferences is assumed to be piecewise linear. We used this to reproduce accurate approximations for some well-known examples of such problems for which closed-form solutions have been derived in the literature. However, we believe that our approach will be particularly useful if it can be utilized for cases where the solution is unknown. This is for example the case when preferences are not know in parametric form, but have to be approximated based on empirical evidence which is based on a limited number of experiments. When the detailed behaviour of the utility function is not known but its general shape is know, our method can be used to calculate optimal investment strategies based on a crude piecewise linear approximation. 

  We also note that it is very easy in our framework to incorporate state-dependent payments which lead to a change of wealth, since such a payment can simply be represented by a horizontal shift of the value function.

 There is a number of extensions of this work that may be interesting. In this paper we restrict ourselves to the case where multiple risk factors may influence our single risky asset, but we do not treat the case where investment in more than one risky asset is possible. Introducing this possibility will introduce more sets of parallel lines when defining the grid that must contain the optimal investment strategy, which clearly complicates the analysis and the design of efficient algorithms. We hope to address this issue in subsequent work.

\bibliography{NoteUdineRefs}

\end{document}